\documentclass[twocolumn, fp]{jpsj3}

\usepackage{txfonts}
\usepackage{amsmath,amssymb,amsfonts,bm}
\usepackage{mathtools}
\usepackage{color}
\usepackage{here}
\usepackage{physics}
\usepackage{braket}

\title{
Microscopic Conversion of Structural Chirality into Electronic Chirality 
\\ in Artificial Chiral Clusters
}

\author{
    Tenchi Endo and 
    Satoru Hayami 
    }
\inst{
Graduate School of Science, Hokkaido University, Sapporo 060-0810, Japan \\
}

\abst{
    We investigate how structural chirality is converted into electronic chirality in a two-dimensional cluster model consisting of a triangular-lattice core and three rod-like extensions under a surface-induced polar field. 
    By evaluating the electric toroidal monopole (ETM) and the Edelstein effect, we show that both quantities exhibit antisymmetric and nonmonotonic dependence 
    on the structural parameter controlling the cluster geometry. 
    While the total density of states is almost unaffected by the structural deformation, the ETM-resolved spectrum changes significantly, indicating that chirality is encoded in the symmetry character of the electronic states rather than in the overall electronic spectrum. 
    Real-space analysis reveals that the induced ETM is concentrated near the vertices and rods, whereas the Edelstein response extends more broadly into the core region. 
    Furthermore, a parameter-decomposition analysis identifies spin--orbit coupling, surface-induced parity mixing, and structural chirality as the essential ingredients for both phenomena. 
    Our results clarify the microscopic connection between structural chirality, electronic chirality, and chiral transport responses in artificial surface nanostructures.
}

\begin{document}
\maketitle

\section{Introduction}

Chirality is a fundamental concept that permeates a wide range of scientific disciplines, including chemistry, biology, and materials science~\cite{naaman2019chiral,cheong2021permutable,cheong2022linking,Togawa2023,Bloom2024,Bousquet2025,Juraschek2025,Zhang2026}.
A chiral object cannot be superposed onto its mirror image by any proper rotation or translation, resulting in two distinct configurations with opposite handedness~\cite{kelvin1894molecular,L.D.Barron_1986_true-chirality,Barron_mol-light-scattering}.
In condensed matter systems, the absence of mirror and inversion symmetries gives rise to a variety of unconventional phenomena, including natural optical activity~\cite{Barron_mol-light-scattering,PhysRevLett.104.163901}, nonlinear responses~\cite{tokura2018nonreciprocal,Sodemann_PhysRevLett.115.216806,li2023current,doi:10.7566/JPSJ.93.043702,Kanda_PhysRevB.111.035130,95kq-fss8,inda2026antiparallel}, electrical magnetochiral effects~\cite{Rikken_PhysRevLett.87.236602,yokouchi2017electrical,Aoki_PhysRevLett.122.057206,rikken2022dielectric}, and the Edelstein effect~\cite{edelstein1990spin,yoda2015current,furukawa2017observation,Furukawa_PhysRevResearch.3.023111}.
Chirality is also closely intertwined with topological magnetism, where the Dzyaloshinskii--Moriya interaction induced by inversion-symmetry breaking stabilizes nontrivial spin textures such as skyrmion and hedgehog crystals~\cite{dzyaloshinsky1958thermodynamic,moriya1960anisotropic,Muhlbauer_2009skyrmion,yu2010real,nagaosa2013topological,Tokura_doi:10.1021/acs.chemrev.0c00297,hayami2024stabilization}. Representative examples include chiral magnets MnSi~\cite{ishikawa1976helical,Muhlbauer_2009skyrmion,Neubauer_PhysRevLett.102.186602} and EuPtSi~\cite{kakihana2018giant,kaneko2019unique,tabata2019magnetic,kakihana2019unique,hayami2021field}.
These studies highlight chirality as a key ingredient of functional and topological states of matter and motivate understanding how structural chirality is encoded into microscopic electronic degrees of freedom.

Recent advances in nano-fabrication technologies have enabled precise manipulation of artificial structures over a wide range of length scales.
Consequently, chiral metamaterials, plasmonic systems, and nano-patterned electronic devices have attracted considerable attention owing to their ability to realize functionalities beyond those of the constituent materials~\cite{PhysRevLett.95.227401, zheludev2010road, plum2015chiral,hentschel2017chiral, wang2017origami, probst2021mechano, PhysRevB.107.155419}.
Furthermore, deformable and reconfigurable architectures provide continuous control of structural parameters, making it possible to directly examine how geometric structures influence physical responses~\cite{Kan2015-ba,kuzyk2017selective, zhao2024mechanically,Zhao2025-vp}.
These developments have established artificial chiral structures as an ideal platform for investigating the microscopic interplay between structural chirality and electronic degrees of freedom.

Despite extensive investigations of chiral materials and phenomena, how structural chirality is encoded into electronic degrees of freedom remains incompletely understood.
In particular, although many chiral functionalities have been observed macroscopically, the microscopic electronic quantity responsible for chirality and its associated responses has yet to be fully clarified~\cite{Miki_PhysRevLett.134.226401, ghyy-mynx}.
From the viewpoint of multipole representation~\cite{kusunose2022generalization, hayami2024unified}
, the electric toroidal monopole (ETM) has recently been proposed as an electronic order parameter for chiral states~\cite{PhysRevB.98.165110,https://doi.org/10.1002/ijch.202200049, hayami2025chirality, kuniyoshi2026theory}.
Since the ETM is odd under spatial inversion and even under time reversal, it shares the same symmetry as chirality and provides a useful descriptor of chiral electronic states in systems ranging from twisted methane molecules~\cite{inda2024quantification} to elemental Te~\cite{Oiwa_PhysRevLett.129.116401, oiwa2025predominant}.
Furthermore, it serves as a symmetry-based order parameter for identifying and characterizing unconventional chiral phase transitions in materials such as Cd$_2$Re$_2$O$_7$\cite{yamaura2002low,Matteo_PhysRevB.96.115156,hiroi2018pyrochlore,Hayami_PhysRevLett.122.147602}, URu$_2$Si$_2$\cite{Kambe_PhysRevB.97.235142,kambe2020symmetry,hayami2023chiral}, and URhSn~\cite{harima2023hidden,kusunose2024configuration,tabata2025successive,ishitobi2026purely}.

Motivated by the growing interest in chiral functionalities and artificial chiral structures, we investigate how structural chirality is encoded into electronic degrees of freedom from a microscopic perspective.
For this purpose, we study a two-dimensional cluster model composed of atomic $s$ and $p$ orbitals in the presence of a surface-induced polar field.
The threefold-symmetric cluster geometry and the polar field cooperate to break all vertical mirror symmetries, thereby generating a finite 
ETM as a measure of electronic chirality.
By continuously tuning a structural parameter that controls the degree of geometric chirality, we systematically examine the evolution of the electronic structure, the ETM, and the Edelstein effect.
Through this analysis, we elucidate how structural chirality is converted into electronic chirality and demonstrate their close correspondence over a wide parameter range.

The remainder of this paper is organized as follows.
In Sect.~\ref{sec:model}, we introduce the two-dimensional cluster model and formulate the structural and electronic chiralities associated with the cluster geometry and the surface-induced polar field.
In Sect.~\ref{sec:result}, we investigate how the ETM and the Edelstein effect evolve with structural chirality by examining their dependence on the structural parameter and their real-space distributions.
We further clarify their microscopic origins through a decomposition analysis for the minimal cluster model.
Finally, Sect.~\ref{sec:sd} is devoted to discussion and concluding remarks.

\section{Model}
\label{sec:model}
\begin{figure}[t]
    \includegraphics[width=\columnwidth]{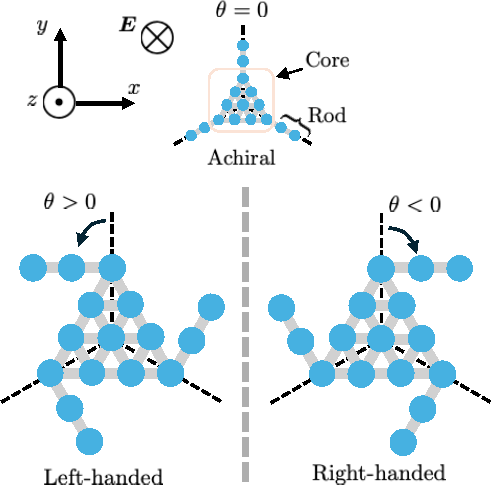}
    \caption{
        Schematic illustration of the cluster geometry and model setup. The cluster consists of a triangular-lattice core and three rod-like extensions attached to its vertices.
        A polar field $\bm{E}$ is present perpendicular to the cluster plane along the $z$ direction.
        The angle $\theta$, measured from the radial direction, controls the structural chirality of the cluster.
        The structural handedness is determined by the combination of $\theta$ and the direction of the polar field.
    }
    \label{fig:model}
\end{figure}

We consider a minimal microscopic model to investigate how structural chirality is encoded into electronic degrees of freedom in artificial surface nanostructures.
As illustrated in Fig.~\ref{fig:model}, the system consists of a two-dimensional cluster with threefold rotational symmetry formed on a surface.
Owing to the broken inversion symmetry associated with the surface environment, an out-of-plane polar field is generated perpendicular to the cluster plane, providing a microscopic source of parity mixing in the electronic states.
The cluster comprises a triangular-lattice core with an equilateral-triangular shape and three rod-like extensions attached to its vertices.
The orientation of the rods relative to the core provides a tunable geometric degree of freedom, enabling continuous control of the structural chirality.
The structural handedness is 
determined by the combination of the rod orientation and the polar-field direction: 
viewed along the polar field, the cluster is classified as right-handed (left-handed) when the rods are rotated clockwise (counterclockwise) with respect to the radial directions.

To describe the electronic structure, we consider atomic $s$ and $p$ orbitals on each site,
which provide the minimal basis for incorporating parity mixing induced by the surface polar field.
The Hamiltonian is given by
\begin{align}
    \mathcal{H}
     & =
    \sum_
    {\langle i, j\rangle}
    \sum_{\alpha, \beta}\sum_{\sigma}
    (t_{ij
    }^{\alpha\beta} c_{i, \sigma}^{\alpha\dagger} c_{j, \sigma}^{\beta} +
    {\rm h.c.})
    +
    \frac{\lambda}{2}\sum_{i}\bm{\sigma}_i\cdot\bm{l}_i
    \nonumber \\&+
    V_{sz}\sum_{i}\sum_{\sigma}
    (c_{i, \sigma}^{s\dagger} c_{i, \sigma}^{z} +
    {\rm h.c.})
    +
    \Delta_{\rm sp}\sum_{i}\sum_{\alpha=x,y,z}
    \sum_{\sigma}
    c_{i, \sigma}^{\alpha\dagger} c_{i, \sigma}^{\alpha}
    \nonumber \\&+
    \Delta_{\rm R}\sum_{i \in {\rm Rod}}
    \sum_{\alpha}\sum_{\sigma}
    c_{i, \sigma}^{
            \alpha \dagger} c_{i, \sigma}^{\alpha} .
\end{align}
Here, $c_{i,\sigma}^{\dagger\alpha}$ ($c_{i,\sigma}^{\alpha}$) is the creation (annihilation) operator for an electron with spin $\sigma$ in orbital $\alpha$ at site $i$, where $\alpha=s, x,y,z$; for notational simplicity, we use $(x,y,z)$ to represent the $(p_x,p_y,p_z)$ orbitals.

The first term describes electron hopping between nearest neighbor sites.
The hopping amplitudes $t_{ij}^{\alpha\beta}$ are expressed using the Slater--Koster parameters $(ss\sigma)$, $(sp\sigma)$, $(pp\sigma)$, and $(pp\pi)$, thereby incorporating the directional character of the $s$ and $p$ orbitals. 
Using bonding angle $\phi$ maesured from the positive 
    $x$-axis, they are given by~\cite{PhysRev.94.1498}
    \begin{align}
        t_{ij}^{ss} & = -(ss\sigma),                              \\
        t_{ij}^{sx} & = -(sp\sigma)\cos\phi,                      \\
        t_{ij}^{sy} & = -(sp\sigma)\sin\phi,                      \\
        t_{ij}^{xy}
        & = t_{ij}^{yx}
        =  (-(pp\sigma)+(pp\pi))\sin\phi\cos\phi,                 \\
        t_{ij}^{xx} & = -(pp\sigma)\cos^2\phi -(pp\pi)\sin^2\phi, \\
        t_{ij}^{yy} & = -(pp\sigma)\sin^2\phi -(pp\pi)\cos^2\phi, \\
        t_{ij}^{zz} & = -(pp\pi).
    \end{align}

The second term represents the atomic spin--orbit coupling (SOC) with strength $\lambda$, where $\bm{\sigma}_i/2$ and $\bm{l}_i$ denote the spin and orbital angular momentum operators, respectively.
The third term describes the on-site hybridization between the $s$ and $p_z$ orbitals induced by the polar field along the out-of-plane direction, which 
explicitly breaks inversion symmetry. 
The fourth term $\Delta_{sp}$ controls the energy splitting between the $s$ and $p$ orbitals, while the last term $\Delta_{\rm R}$ modifies the onsite energy in the rod region, allowing the rods to have an electronic environment different from that of the central triangular cluster.

To characterize the electronic chirality, we evaluate the ETM, which has been proposed as a microscopic order parameter for chiral electronic states~\cite{PhysRevB.98.165110,https://doi.org/10.1002/ijch.202200049}.
The operator expression of total atomic ETM is defined as
\begin{align}
    G_0^{\rm total}
            & =
    \sum_i (G_{0})_i.         \\
    (G_0)_i & =\sum_{a=x,y,z}
    \sigma_{a,i}\otimes T_{a,i},
\end{align}
where $\sigma_{a,i}$ and $T_{a,i}$ denote the spin and magnetic toroidal dipole operators at site $i$, respectively.
Here, the magnetic toroidal dipole operator at site $i$ is defined as~\cite{hayami2018microscopic}
\begin{align}
        &
    T_{x}=\frac{1}{3\sqrt{3}}
    \left(
    \begin{array}{cccc}
            0  & i & 0 & 0 \\
            -i & 0 & 0 & 0 \\
            0  & 0 & 0 & 0 \\
            0  & 0 & 0 & 0 \\
        \end{array}
    \right),
    \cr &
    T_{y}=\frac{1}{3\sqrt{3}}
    \left(
    \begin{array}{cccc}
            0  & 0 & i & 0 \\
            0  & 0 & 0 & 0 \\
            -i & 0 & 0 & 0 \\
            0  & 0 & 0 & 0 \\
        \end{array}
    \right),
    \cr &
    T_{z}=\frac{1}{3\sqrt{3}}
    \left(
    \begin{array}{cccc}
            0  & 0 & 0 & i \\
            0  & 0 & 0 & 0 \\
            0  & 0 & 0 & 0 \\
            -i & 0 & 0 & 0 \\
        \end{array}
    \right),
\end{align}
for the basis $( s, p_x, p_y, p_z )$.
The magnetic toroidal dipole represents an electronic degree of freedom associated with imaginary $s$--$p$ hybridization.
Because $\bm{\sigma}$ is an axial vector while $\bm{T}=(T_x, T_y, T_z)$ is a polar vector, the ETM $G_0\propto\bm{\sigma}\cdot\bm{T}$ is a pseudoscalar quantity.
As a consequence, it changes sign under spatial inversion but remains invariant under time reversal, exactly matching the symmetry of chirality.
The ETM therefore provides a natural microscopic descriptor of electronic chirality.

In addition to the total ETM, we separately evaluate the contributions from the central triangular region and the rods to clarify their respective roles in generating electronic chirality.
The corresponding quantities are defined as
\begin{align}
    G_0^{\rm core}
     & =
    \sum_{i\in {\rm Core}}
    (G_0)_i,
    \\
    G_0^{\rm rod}
     & =
    \sum_{i\in {\rm Rod}}
    (G_0)_i,
\end{align}
where the summations are taken over the sites belonging to the triangular core and the rod regions, respectively. By construction,
\begin{align}
    G_0^{\rm total}
    =
    G_0^{\rm core}
    +
    G_0^{\rm rod}.
\end{align}
In the following, the expectation value of the ETM is used to quantify electronic chirality.

To investigate the transport response associated with electronic chirality, we calculate the Edelstein effect within linear response theory.
The response tensor is given by
\begin{align}
    \chi_{a, b} 
    =
    -\frac{{\rm i}}{V\sigma_{b,b}}
    \sum_{m, n}
    \frac{f(\varepsilon_m)-f(\varepsilon_n)}{\varepsilon_m- \varepsilon_n} \frac{\braket{m| S_a|n}\braket{n|J_b|m}}{\varepsilon_m- \varepsilon_n 
    + {\rm i} \delta},
\end{align}
with
\begin{align}
    \sigma_{a, b}(\omega)
    =
    -\frac{{\rm i}}{V}
    \sum_{m, n}
    \frac{f(\varepsilon_m)-f(\varepsilon_n)}{\varepsilon_m- \varepsilon_n} \frac{\braket{m| J_a|n}\braket{n|J_b|m}}{\varepsilon_m- \varepsilon_n 
    + {\rm i} \delta}. 
\end{align}
Here, $|m\rangle$, $\varepsilon_m$, and $f(\varepsilon_m)$ denote the eigenstate, eigenenergy, and Fermi--Dirac distribution function for the $m$th band, respectively, and $V$ is size of the system.
The operators $S_a$ and $J_a$ represent the spin magnetization and charge-current operators along the $a$ direction.
The quantity $\chi_{a,b}$ measures the current-induced spin polarization, while $\sigma_{a,b}$ is the electrical conductivity tensor.
In addition, $\hbar$ is the reduced Planck constant and $\delta$ is a phenomenological broadening parameter that accounts for finite phenomenological quasiparticle lifetimes.
In the present time-reversal-symmetric system, we focus on the intraband contribution to the dissipative response~\cite{Yanase2014-na}.
By comparing the ETM and Edelstein responses over a wide range of structural parameters, we clarify how structural chirality is converted into electronic chirality and chiral transport responses.

\section{Result}
\label{sec:result}
In the following numerical calculations, we set the Slater--Koster parameters as
\begin{align}
    ((ss\sigma),(sp\sigma),(pp\sigma),(pp\pi))
    =
    (1,-0.3,-0.6,0.25),
\end{align}
and use
\begin{align}
    \lambda=0.3,\quad
    V_{sz}=-0.2,\quad
    (\Delta_{sp}, \Delta_{\rm R})=(10,5).
\end{align}
We also set $e=\hbar=\mu_{\rm B}=1$, the $g$ factor to $g=2$, and the phenomenological broadening parameter to $\delta=0.001$.
Unless otherwise specified, the cluster used in the calculations consists of a triangular-lattice core with 64 sites along each edge and three rods of length 5 attached to its vertices; the total number of sites is given by $2080+5\times 3 = 2095$. 
    The electron filling per site is fixed at 10430/2095 $\sim$ 4.98 so that the chemical potential is located at the midpoint between the highest occupied level corresponding to half filling of the core $p$ orbitals and the lowest unoccupied level above it. 

\subsection{
    Structural-parameter dependence of the ETM and Edelstein effect}

\begin{figure}[t]
    \includegraphics[width=\columnwidth]{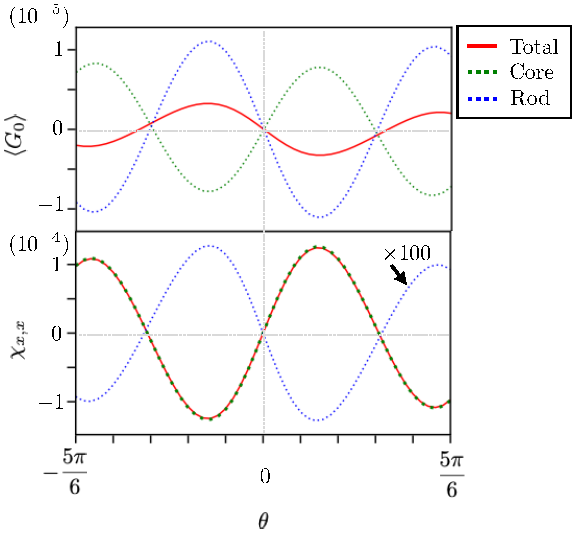}
    \caption{
        Structural-parameter dependence of the electric toroidal monopole (ETM) and the Edelstein response
        $\chi_{x, x}$. 
        The upper and lower panels show the ETM and  $\chi_{x, x}$, respectively.
        Red, green, and blue curves represent the total, core, and rod contributions.
        The rod contribution to
        $\chi_{x, x}$ 
        is multiplied by 100 for visibility.
        Both quantities change sign upon reversing $\theta$, reflecting the reversal of electronic chirality associated with the structural handedness.
    }
    \label{f2}
\end{figure}

We first examine how structural chirality is reflected in the electronic chirality and the current-induced spin response.
Figure~\ref{f2} shows the $\theta$ dependence of the ETM and the Edelstein response  $\chi_{x, x}$, together with their decompositions into the core and rod contributions.
Both the ETM and $\chi_{x, x}$ exhibit antisymmetric behavior with respect to $\theta$, reflecting the reversal of electronic chirality upon reversing the structural handedness.
In the achiral configuration at $\theta=0$, the additional mirror symmetry suppresses the ETM and the Edelstein response.
Once the rods are tilted away from the radial direction, the mirror symmetries are broken and finite values appear.

A notable feature is that the ETM and the Edelstein response do not increase monotonically with the magnitude of the structural distortion.
Instead, they show nontrivial extrema as functions of $\theta$, indicating that the electronic chirality is not determined solely by the degree of geometric deformation.
This behavior demonstrates that the conversion from structural chirality to electronic chirality depends sensitively on the underlying electronic structure.

The decomposition into the core and rod contributions further clarifies the role of each region.
Although the ETM and the Edelstein response exhibit similar $\theta$ dependences, their spatial decompositions reveal different characteristics.
For the ETM, the contributions from the core and rods are comparable in magnitude despite the fact that the rod region contains only a small fraction of the total sites. 
This result indicates that the ETM is strongly concentrated around the rods and their neighboring regions, where the structural chirality is introduced most directly.
In contrast, the rod contribution to the Edelstein response is much smaller than the core contribution, indicating that the overall response is dominated by the electronic states in the core region.
These results suggest that the electronic chirality generated locally near the rods is transferred to the core through orbital hybridization. 
As a consequence, the structurally induced ETM influences the electronic states in the much larger core region and gives rise to a chiral transport response dominated by the core contribution. 
This finding demonstrates how a local structural distortion can generate a macroscopic chiral response through the propagation of electronic chirality.

\subsection{
    Density of states and ETM-resolved electronic states}
\begin{figure}[t]
    \includegraphics[width=\linewidth]{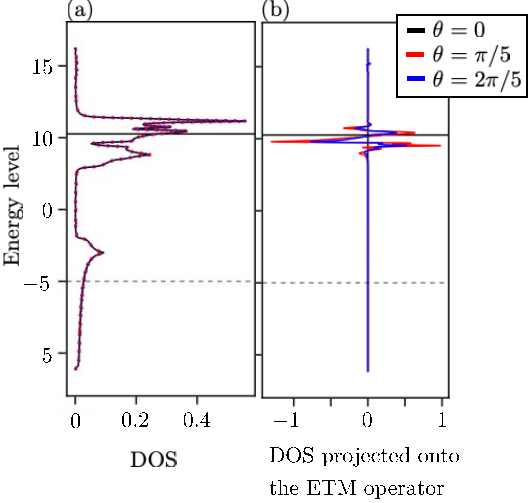}
    \caption{
        (a) Density of states (DOS) and (b) ETM-resolved 
        DOS for representative cluster geometries at $\theta=0$, $\pi/5$, and $2\pi/5$. 
        For better visibility, the values are divided by $10^{-8}$. 
        The horizontal solid line indicates the Fermi level at $\theta=\pi/5$. 
        In (b), DOS is projected onto the ETM operator. 
        Although the total DOS is almost unchanged by the structural deformation, the ETM-resolved spectrum becomes finite.
    }
    \label{f3}
\end{figure}

To clarify whether structural chirality 
modifies the overall electronic spectrum, we calculate the density of states (DOS) for representative values of $\theta$.
Figure~\ref{f3}(a) shows the DOS for the 
achiral structure with $\theta=0$ and the chiral structures with $\theta=\pi/5$ and $2\pi/5$.
The three curves are almost indistinguishable.
This weak dependence on $\theta$ is natural because the structural chirality is introduced by modifying only the rod geometry, while the electronic spectrum is dominated by the much larger triangular core. Therefore, the overall DOS remains nearly unchanged even though the structural chirality is substantially varied.

In contrast, the ETM-resolved DOS shown in Fig.~\ref{f3}(b) exhibits a strong dependence on $\theta$.
Although the total DOS is almost insensitive to the structural deformation, finite ETM-resolved spectral weight emerges in the chiral cluster 
($\theta \neq 0$), whereas it vanishes in the 
achiral cluster ($\theta = 0$) because of the remaining mirror symmetries.
This contrast demonstrates that structural chirality is primarily encoded in the symmetry character of the electronic wave functions rather than in the overall distribution of energy levels.
These results indicate that the ETM can reveal chiral characteristics of the electronic states that are hidden in conventional spectral quantities such as the DOS.

\subsection{
    Real-space distributions of the ETM and Edelstein effect}

\begin{figure*}[t]
    \centering
    \includegraphics[width=\linewidth]{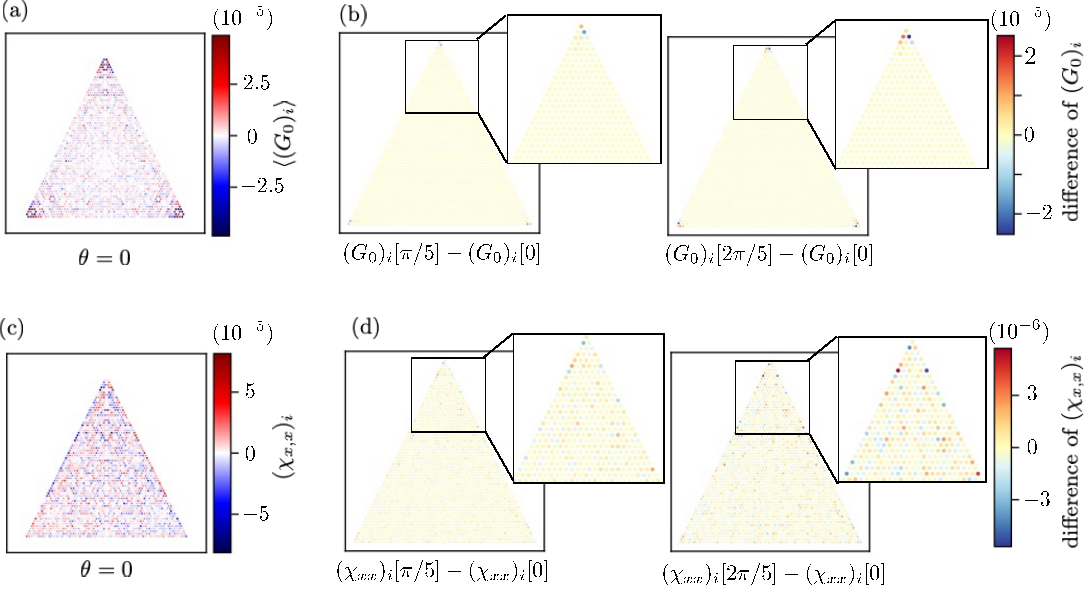}
    \caption{
        Real-space distributions of the site-resolved ETM and Edelstein response.
        Panels (a) and (c) show the onsite ETM and Edelstein response for the achiral cluster.
        Panels (b) and (d) show the differences from the achiral cluster for chiral clusters with (left panel) $\theta=\pi/5$ and (right panel) $2\pi/5$.
        We denote the onsite ETM and Edelstein response at site $i$ in the cluster with angle $\theta$ by $(G_{0})_i[\theta]$ and $(\chi_{x,x})_i[\theta]$, for simplicity.
        The difference maps highlight the components induced by structural chirality.
        }
    \label{f4}
\end{figure*}

We next examine the spatial structures of the electronic chirality and the Edelstein response.
Figures~\ref{f4}(a) and \ref{f4}(c) show the real-space distributions of the onsite ETM and 
site-resolved Edelstein response, respectively, for the 
achiral cluster with $\theta=0$.
The onsite Edelstein response is calculated using onsite spin magnetization operator matrix $(S_{a})_i$ along the $a$ axis as follows 
\begin{align}
    (\chi_{a, b})_i
    =
    -\frac{{\rm i}}{V\sigma_{b,b}}
    \sum_{m, n}
    \frac{f(\varepsilon_m)-f(\varepsilon_n)}{\varepsilon_m- \varepsilon_n} \frac{\braket{m| (S_a)_i|n}\braket{n|J_b|m}}{\varepsilon_m- \varepsilon_n 
    + {\rm i} \delta}.
\end{align}
With this definition, 
the total spin magnetization operator matrix and Edelstein response in the whole system satisfy 
\begin{align}
  S_{a}
  & = 
  \sum_{i} (S_{a})_i 
  \\
  \chi_{a, b}
  & = 
  \sum_{i} (\chi_{a, b})_i. 
\end{align}

In both quantities, sizable local contributions appear mainly near the cluster boundary.
Nevertheless, the total ETM vanishes because the remaining mirror symmetries enforce a cancellation among different regions of the cluster.

To identify the components induced by structural chirality, we compare the chiral clusters with the  
achiral one.
Figures~\ref{f4}(b) and \ref{f4}(d) show the differences in the onsite ETM and Edelstein response, respectively, between the chiral and 
achiral clusters.
As seen in Fig.~\ref{f4}(b), the structurally induced ETM is strongly concentrated near the vertices of the triangular core and the regions adjacent to the rods.
This indicates that the local mirror-symmetry breaking introduced by the rod tilting is most directly encoded into the electronic chirality near the corners of the cluster.

In contrast, the changes in the Edelstein response shown in Fig.~\ref{f4}(d) exhibit a broader spatial distribution.
Although enhanced contributions are also observed near the vertices and edges, finite changes extend further into the interior of the core, particularly for the larger structural distortion with $\theta=2\pi/5$.
This difference suggests that the ETM and the Edelstein effect, while closely related by symmetry, probe different aspects of the electronic states: the ETM primarily reflects local electronic chirality, whereas the Edelstein response is influenced by current-carrying states that are spatially more extended.

These real-space distributions provide microscopic insight into how structural chirality is transferred from the rods to the electronic states of the entire cluster.
Although the rods act as the primary geometric source of chirality, the induced ETM and Edelstein response are not confined to the rod region.
Instead, they propagate into the triangular core through electronic hybridization, giving rise to chiral electronic states and transport responses over a much broader spatial region.
This behavior explains why the Edelstein response in Fig.~\ref{f2} is dominated by the core contribution even though the structural chirality originates from the rods. 
In other words, the rod-induced electronic chirality is transmitted to the much larger core region, where it generates the dominant contribution to the chiral transport response.

\subsection{Parameter decomposition of the ETM and Edelstein effect}
Finally, we analyze the microscopic origin of the ETM and the Edelstein effect through a parameter-decomposition analysis~\cite{doi:10.7566/JPSJ.91.014701}.
This method has been widely employed to identify the minimal set of microscopic parameters required for emergent response functions, such as magnetoelectric effects~\cite{Yatsushiro_PhysRevB.105.155157} and nonlinear conductivities~\cite{Hayami_PhysRevB.106.014420, Kirikoshi_PhysRevB.107.155109}.
Applying the same framework to the present model enables us to clarify which microscopic ingredients are responsible for electronic chirality and the resulting Edelstein response.

For this purpose, we evaluate the model-dependent quantities
\begin{align}
    \Gamma^i
     & =
    {\rm Tr} \qty(
    G_0 H^i
    ),
    \\
    \Gamma^ {ij}_{a, b}
     & =
    {\rm Tr} \qty(
    S_a H^i J_b H^j
    ),
\end{align}
where $i$ and $j$ are positive integers.
$H^i$ denotes the $i$th power of the Hamiltonian matrix.
The former characterizes the parameter dependence of the ETM, while the latter represents that of the Edelstein response.
We perform this decomposition for the minimal cluster consisting of a triangular core with three sites and one site in each rod, which allows us to extract the essential parameter combinations analytically.

For the ETM, the leading independent contributions for $i\geq 7$ are expressed as
\begin{align}
    \Gamma^i
     & =
    \Gamma^i_{1} +
    \Gamma^i_{2} +
    \Gamma^i_{3} +
    \Gamma^i_{4},
    \\
    \Gamma^i_{1}
     & \propto
    \lambda V_{sz} (sp\sigma)^2(pp\pi)\sin \theta,
    \\
    \Gamma^i_{2}
     & \propto
    \lambda V_{sz} (sp\sigma)^2\Delta_{\rm R}\sin \theta,
    \\
    \Gamma^i_{3}
     & \propto
    \lambda V_{sz} (pp\pi)\Delta_{\rm R}\sin \theta,
    \\
    \Gamma^i_{4}
     & \propto
    \lambda V_{sz} (ss\sigma)(pp\sigma)^2\sin \theta,
\end{align}
where terms that are not essential for the following symmetry discussion are omitted for simplicity.
The displayed terms show that the ETM requires the simultaneous presence of SOC $\lambda$, surface-induced $s$-$p_z$ hybridization $V_{sz}$, and structural chirality represented by odd functions of $\theta$.

This parameter dependence has a clear physical meaning.
The factor $V_{sz}$ encodes the inversion-symmetry breaking induced by the surface polar field, while $\lambda$ converts the geometric chirality into a spin-dependent electronic chirality.
The factors involving the Slater--Koster parameters describe orbital hopping processes that transmit the geometric deformation to the electronic wave functions.
The appearance of $\sin\theta$ reflects the odd nature of chirality under reversal of the structural handedness.
Thus, the ETM emerges only when the surface polarity, SOC, and chiral geometry cooperate.

A similar conclusion is obtained for the Edelstein response.
The decomposition of $\Gamma^{ij}_{a,b}$ gives finite contributions for combinations such as $(i,j)=(2,5),(3,4)\cdots$.
For the longitudinal component considered above, the common leading factor is
\begin{align}
    \Gamma^ {ij}_{x, x}
     & \propto
    \lambda^2 V_{sz} (sp\sigma)\sin \theta,
\end{align}
again up to terms omitted for clarity.
This expression demonstrates that the Edelstein response is governed by the same essential ingredients as the ETM: SOC, surface-induced parity mixing, and structural chirality.

The parameter decomposition therefore provides a microscopic explanation for the close correspondence between the ETM and the Edelstein effect observed in Fig.~\ref{f2}.
Both quantities originate from the same symmetry-breaking processes and share the characteristic factors $\lambda$, $V_{sz}$ and $\sin\theta$. 
These results indicate that the ETM and the Edelstein effect are governed by common microscopic ingredients associated with structural chirality, SOC, and parity mixing. 
At the same time, the parameter decomposition reveals a subtle difference between the two quantities: 
the leading contribution to the Edelstein response necessarily contains the factor $\lambda^2$, 
whereas the ETM rather depends on the first order of $\lambda$ and can  
arise from other hopping processes. 
This finding further also suggests that the $s$--$p$ hybridization hopping represented by $(sp\sigma)$ may play an important role in transmitting the chirality near the rods to the electronic states in the core region, thereby contributing to the emergence of a macroscopic chiral transport response.

\section{Summary}
\label{sec:sd}

We have investigated how structural chirality is encoded into microscopic electronic degrees of freedom in a two-dimensional chiral cluster model.
The model consists of a triangular-lattice core and three rod-like extensions under a surface-induced polar field.
By tilting the rods away from the radial directions, all vertical mirror symmetries are broken, leading to a structurally chiral cluster.
Within an atomic $s$-$p$ orbital model including SOC and surface-induced $s$-$p_z$ hybridization, we evaluated the ETM and the Edelstein effect as microscopic measures of electronic chirality and its associated response.

We found that both the ETM and the Edelstein response exhibit antisymmetric behavior with respect to the structural parameter $\theta$, reflecting the reversal of electronic chirality upon reversing the structural handedness.
Their $\theta$ dependence is, however, nonmonotonic, indicating that electronic chirality is not determined solely by the magnitude of geometric deformation.
This result demonstrates that the conversion from structural chirality to electronic chirality is governed by the underlying electronic structure.

The DOS analysis showed that the total DOS is almost unchanged by the structural deformation, mainly because the rods occupy only a small fraction of the whole cluster compared with the triangular core.
In contrast, the ETM-resolved DOS exhibits a pronounced dependence on $\theta$.
This contrast indicates that structural chirality is encoded not through a large reconstruction of the energy spectrum, but through a change in the symmetry character of the electronic wave functions. 
Thus, the ETM provides a sensitive probe of electronic chirality that is invisible in conventional spectral quantities.

The real-space distributions further clarified how the structural chirality introduced by the rods is transferred to the cluster.
The structurally induced ETM is mainly localized near the vertices and the regions adjacent to the rods, where the local mirror-symmetry breaking is most pronounced.
On the other hand, the change in the Edelstein response extends more broadly into the triangular core. This difference indicates that the ETM mainly reflects local electronic chirality, whereas the Edelstein response is influenced by spatially extended current-carrying states.
Nevertheless, the close correspondence between their structural-parameter dependence 
suggests a strong correlation between the ETM and the Edelstein effect, both of which originate from the same symmetry-breaking mechanisms.

The parameter--decomposition analysis revealed the microscopic ingredients necessary for generating the ETM and the Edelstein effect.
Both quantities contains the same key microscopic ingredients, $\lambda$, $V_{sz}$ and $\sin\theta$,
indicating that SOC, surface-induced parity mixing, and structural chirality may constitute the common microscopic ingredients underlying both the ETM and the Edelstein effect. 
This result provides a microscopic explanation for the strong correlation between electronic chirality and chiral transport responses in the present system.

In summary, our results demonstrate that a small structural deformation introduced by rod-like components can induce chiral electronic degrees of freedom in a much larger core region.
The present study clarifies the microscopic link between structural chirality, electronic chirality, and chiral response phenomena in artificial surface nanostructures.
These findings provide a basis for designing chiral electronic functionalities through controlled geometric structures.

\begin{acknowledgments}
    This research was supported by JSPS KAKENHI (Grant Nos. JP22H00101 and JP23H04869),
    JST CREST (Grant No.~JPMJCR23O4), and JST FOREST (Grant No.~JPMJFR2366).
\end{acknowledgments}

\bibliography{jpsj_ref.bib}
\bibliographystyle{jpsj}
\end{document}